\documentclass[10pt, twocolumn]{article}

\usepackage[vlined, ruled, linesnumbered, noend]{algorithm2e}
\usepackage{lhelp}
\usepackage{hyperref}
\usepackage{graphicx}
\usepackage{amsmath}
\usepackage{amssymb}
\usepackage{amsfonts,latexsym,amssymb}
\usepackage{authblk}

\def\KK {\ensuremath{\mathbb{K}}}
\def\N {\ensuremath{\mathbb{N}}}

\newcommand{\MMM}{MMM}
\renewcommand{\P}{\mbox{\sf P}}
\newcommand{\K}{\mbox{\sf K}}

\newtheorem{Corollary}{Corollary}
\newtheorem{Theorem}{Theorem}

\newif\ifshow
\showfalse

\newif\iflongversion
\longversionfalse

\begin{document}

\title{A Many-core Machine Model for Designing Algorithms with Minimum Parallelism Overheads}

\author{Sardar Anisul Haque}
\affil{\href{mailto:shaque4@csd.uwo.ca}{shaque4@csd.uwo.ca} \\
Department of Computer Science, 
University of Western Ontario
}

\author{Marc Moreno Maza}
\affil{\href{mailto:moreno@csd.uwo.ca}{moreno@csd.uwo.ca} \\
Department of Computer Science, 
University of Western Ontario
}

\author{Ning Xie}
\affil{\href{mailto:nxie6@csd.uwo.ca}{nxie6@csd.uwo.ca} \\
Department of Computer Science, 
University of Western Ontario
}

\date{\vspace{-5ex}}

\maketitle

\begin{abstract}
 We present a model of multithreaded computation, combining
fork-join and single-instruction-multiple-data parallelisms, with
an emphasis on estimating parallelism overheads of programs
written for modern many-core architectures.  We establish a
Graham-Brent theorem for this model so as to estimate execution
time of programs running on a given number of streaming
multiprocessors.  We evaluate the benefits of our model with
four fundamental algorithms from scientific computing.  In each
case, our model is used to minimize parallelism overheads
by determining an appropriate value range for a given program
parameter; moreover experimentation confirms the model's prediction. 

\end{abstract}


\section{Introduction}

Designing efficient algorithms targeting implementation
on hardware acceleration technologies (multi-core processors, 
graphics processing units (GPUs), field-programmable gate arrays)
creates major challenges for computer scientists.
A first difficulty is to define models of computations
retaining the computer hardware characteristics
that have a dominant impact on program performance.
Therefore, in addition to specify the appropriate complexity measures
for the algorithms to be analyzed,
those models must consider the relevant parameters characterizing the
abstract machine executing those algorithms.
A second difficulty is, 
for a given model of computations,
to combine its complexity measures so as
to determine the ``best'' algorithm among different algorithmic solutions
to a given problem.

In the fork-join concurrency model~\cite{DBLP:journals/siamcomp/BlumofeL98}
two complexity measures, the work $T_1$ and the span $T_{\infty}$, 
and one machine parameter, the number ${\P}$ of processors,
can be combined in results like the
Graham-Brent theorem~\cite{DBLP:journals/siamcomp/BlumofeL98,Graham69boundson}
or the Blumofe-Leiserson theorem (Theorems 13 \&  14 
\iflongversion in~\cite{10.1109/SFCS.1994.365680})
\else in~\cite{DBLP:journals/jacm/BlumofeL99}) \fi 
in order to compare algorithm running time estimates.
We recall that the Graham-Brent theorem states that the running time
$T_{\P}$ on ${\P}$ processors satisfies $T_{\P} \leq T_1/{\P} + T_{\infty}$.
A refinement of this theorem supports
the implementation (on multi-core architectures)
of the parallel performance analyzer {\tt Cilkview}~\cite{DBLP:conf/spaa/HeLL10}.
In this context, the running time
$T_{\P}$ is bounded in expectation by $T_1/{\P} + 2 {\delta} \widehat{T_{\infty}}$,
where ${\delta}$ is a constant (called the {\em span coefficient})
and $\widehat{T_{\infty}}$ is the burdened span.

With the pervasive ubiquity of many-core processors, in particular GPUs, 
it is desirable for models of computations 
to combine explicitly both 
task-based parallelism and data-based parallelism.
In fact, popular concurrency platforms 
({CilkPlus}~\cite{DBLP:journals/tjs/Leiserson10,robison2013composable},
 {CUDA}~\cite{Nickolls:2008:SPP:1365490.1365500,nvidia2009nvidia} and
 {OpenCL}~\cite{stone2010opencl}) 
offer both forms of parallelism, with language constructs specific
to each case.
Meanwhile, classical models of parallel computations,
like the fork-join concurrency model or the 
PRAM model~\cite{DBLP:journals/siamcomp/StockmeyerV84,Gibbons:1989:MPP:72935.72953},
do not distinguish between task-based and data-based parallelism,
which is too simplistic for analyzing algorithms
targeting the above concurrency platforms.
In addition, the PRAM model fails to retain
important features of actual computers related to memory traffic,
such as cache \iflongversion complexity~(\cite{FrigoLeisersonRandallCilk51998,Frigo:2006:CCM:1148109.1148157}).
This latter notion has been proved to be very useful 
on single-core and multi-core multiprocessors.
\else complexity~\cite{DBLP:conf/focs/FrigoLPR99}.
\fi

An attempt to integrate memory contention into the PRAM model 
has been made with the QRQW (Queue Read Queue Write) PRAM, defined in~\cite{GMR1998}
by Gibbons, Matias and Ramachandran.
The authors also enhance the Graham-Brent theorem.
However, they conflate in a single quantity time spent in arithmetic
operations and time spent in read/write accesses.
We believe that this unification is not appropriate
for recent many-core processors, such as GPUs,
for which the ratio between one global memory read/write access
and one floating point operation  can be in the 100's.

In a recent paper, Ma, Agrawal and Chamberlain~\cite{MAC2012} 
introduce the TMM (Threaded Many-core Memory) model which
retains many important characteristics of GPU-type architectures,
including several machine parameters such as throughput and 
coalesced granularity. Moreover, TMM analysis can rank algorithms 
from slow to fast, given those machine parameters, 
while their running time estimate on ${\P}$ cores 
is not based on the Graham-Brent theorem.

Many works, such as~\cite{MC2012,WMS2007},  targeting code optimization
and performance prediction of GPU programs
are related to our work. However, these papers
do not define an abstract model in support of the
analysis of algorithms.

In this 
\iflongversion
chapter,
\else
paper,
\fi
 we propose a many-core machine model ({\MMM})
which aims at minimizing  parallelism overheads of 
algorithms targeting implementation on GPUs.
In the design of this model, we insist on the following features:
\begin{itemize}
\item[-] {\em  Two-level DAG programs.}
Defined in Section~\ref{sect:MMM}, this aspect 
captures the two levels of parallelism 
(fork-join and SIMD) of heterogeneous programs
(like a CilkPlus program using 
 {\tt \#pragma simd}~\cite{robison2013composable} or a CUDA program with
the so-called dynamic parallelism~\cite{nvidia2009nvidia}).
\item[-] {\em  Parallelism overhead.} We introduce this complexity
measure in Section~\ref{sect:MMMmeasures} with the objective
of analyzing communication and synchronization costs.
\item[-] {\em  A Graham-Brent theorem.} We combine three 
complexity measures (work, span and parallelism overhead) and
two machine parameters (size of local memory and data transfer
throughput) in order to estimate the running time of an {\MMM} program
on $P$ streaming multiprocessors.
This result is Theorem~\ref{thrm:GrahamBrendt} in Section~\ref{sect:GrahamBrent}.
\end{itemize}
Our proposed model extends both
the fork-join concurrency model and PRAM-based models,
with an emphasis on parallelism overheads resulting
from communication and synchronization costs.

We sketch below how, in practice, we use this model in order to minimize
parallelism overheads of programs targeting GPUs.
Consider an {\MMM} program ${\cal P}$, that is, 
an algorithm expressed in the {\MMM} model.
Assume that a program parameter $s$
(like the number of threads per thread-block
or the amount of data transfer per thread-block)
can be arbitrarily chosen within some range ${\cal S}$ while
preserving the specifications of ${\cal P}$.
Let $s_0$  be a particular value of $s$ which
corresponds to an instance ${\cal P}_0$ of ${\cal P}$ that, 
in practice, can be seen as a naive (or simply initial) version of 
the algorithm. 

Assume that, when $s$ varies within ${\cal S}$, the work, say $W_{\cal P}(s)$,
does not vary much,
that is, $W_{\cal P}(s_0) / W_{\cal P}(s)$ $\in \Theta(1)$ hold.
Assume also that the
parallelism overhead $O_{\cal P}(s)$ varies more substantially,
say $O_{\cal P}(s_0) / O_{\cal P}(s) \in {\Theta}(s - s_0)$.
Then, we determine a value $s_{\rm min} \in {\cal S}$ 
maximizing the ratio $O_{\cal P}(s_0) / O_{\cal P}(s)$.
Next, we use our version of Graham-Brent theorem  
(more precisely, we use Corollary~\ref{coro:Ning})
to check that the upper bound for the running time 
of ${\cal P}(s_{\rm min})$ is less than that of ${\cal P}(s_o)$.
If this holds, we view ${\cal P}(s_{\rm min})$ 
as a solution of our problem of algorithm optimization (in terms of 
parallelism overheads).

To demonstrate and evaluate the benefits of our model, 
and the above optimization strategy, 
we applied them successfully to four fundamental algorithms 
in scientific computing, see 
\iflongversion
Chapters~\ref{chap:plainGCDandDivision} and~\ref{chap:plainMultiplication}
\else
Sections~\ref{sect:division} to~\ref{sect:gcd}.
\fi 
These four applications are the univariate 
polynomial division and multiplication, radix sort
and the Euclidean algorithm.
Each of them satisfies the hypotheses of the above optimization strategy.
Our theoretical analysis,
for three of these applications,  
has led to an optimized implementation
reported in~\cite{HM2012} and publicly available\footnote{
Our algorithms are implemented in CUDA and publicly available 
with benchmarking scripts from~\url{http://www.cumodp.org/}.}.
while, for the other application, this analysis has explained  a posteriori
the experimental observations of~\cite{SHG2009}.

\iflongversion
Finally, we observe that, in our model, the Euclidean algorithm reaches 
the running estimates predicted
by the Systolic VLSI Array Model~\cite{DBLP:journals/tc/BrentK84}.
At the same time, the CUDA code implementing 
the Euclidean algorithm runs within the same estimate 
analyzed by our model for input polynomials 
with degree up to 100,000, as reported in~\cite{HM2012}.
\fi

\section{A many-core machine model}
\label{sect:MMM}

The model of parallel computations presented in this paper
aims at capturing communication and synchronization
overheads of programs written for modern many-core architectures.
One of our objectives is to optimize algorithms
by techniques like reducing redundant memory accesses.
The reason for this optimization 
is  the fact that, on actual GPUs, global memory latency 
is approximately 400 to 800 clock cycles, while local memory latency is 
only a few clock cycles. 
This memory latency difference, when not properly taken into account,
may have a dramatically negative impact on program performance.
As mentioned in the introduction, 
this hardware feature of GPUs cannot be
captured by the well-studied PRAM model.
Indeed, any memory access, as well as any integer arithmetic 
operation, is performed in {\em unit time} on a PRAM machine.

This latter, as well as other limitations, have motivated 
variants of the PRAM model, including our work.
Another motivation, mentioned above, is the 
fact that popular concurrency platforms offer 
task-based and data-based parallelisms,
with language constructs specific to each case.

As specified in Sections~\ref{sect:MMMcharacteristics}
and \ref{sect:MMMprograms},
our many-core machine model ({\MMM}) retains many of the
characteristics of modern GPU architectures 
programming models like CUDA or OpenCL. 
However, in order to support algorithm analysis
with an emphasis on parallelism overheads,
as defined in Section~\ref{sect:MMMmeasures} and
\ref{sect:GrahamBrent},
{\MMM} abstract machines admit a few simplifications
and limitations with respect to actual many-core devices.
\iflongversion
In Section~\ref{sect:MMMjustification}, 
we justify how more general programming models 
can be reduced to ours.
\fi

\iflongversion
\subsection{Characteristics of the many-core abstract machine}
\label{sect:MMMcharacteristics}

  \begin{figure}[htb]
    \begin{center}
    \includegraphics[scale=0.45]{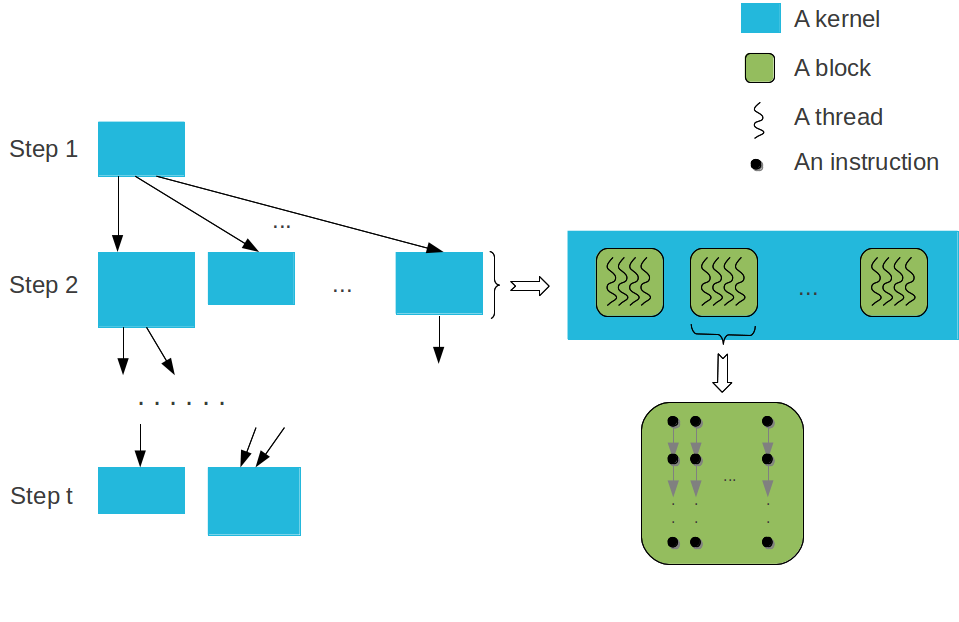}
    \end{center}
    \caption{Overview of a many-core machine program}
    \label{fig:MMMprogram}
  \end{figure}

\else
\subsection{Characteristics of the abstract many-core machines}
\label{sect:MMMcharacteristics}

  \begin{figure}[htb]
    \begin{center}
    \includegraphics[trim=0 40 0 0,clip,scale=0.3]{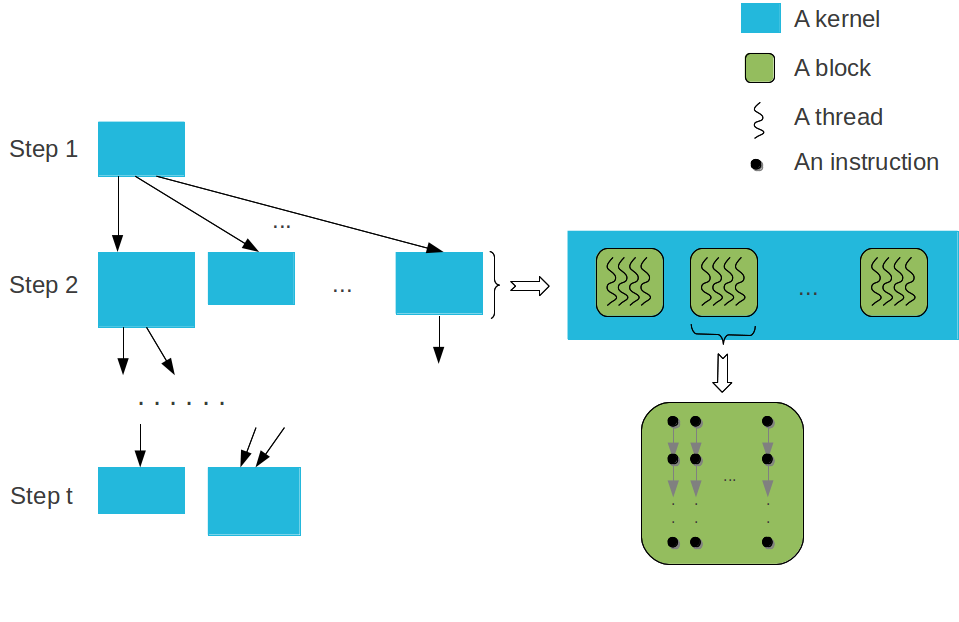}
    \end{center}
    \caption{Overview of a many-core machine program}
    \label{fig:MMMprogram}
  \end{figure}

\fi

\smallskip\noindent{\small \bf Architecture.}
An {\MMM} abstract machine possesses an unbounded number of 
{\em streaming multiprocessors} (SMs) which are all identical.
Each SM has a finite number of processing cores and a fixed-size local memory.
An {\MMM} machine has a two-level memory hierarchy, comprising an 
unbounded global memory with high latency and low throughput 
while SMs local memories have 
low latency and high throughput. 

\smallskip\noindent{\small \bf Programs.}
An {\em {\MMM} program} is a directed acyclic graph (DAG) whose vertices
are kernels (defined hereafter) and edges indicate serial dependencies,
similarly to the instruction stream DAGs of the fork-join 
multithreaded concurrency model.
A {\em kernel} is an SIMD (single instruction, multiple data)
program capable of branches and 
decomposed into a number of thread-blocks. 
Each {\em thread-block} is executed by a single SM
and each SM executes a single thread-block at a time. 
Similarly to a CUDA program, an {\MMM} program specifies for each kernel 
the number of thread-blocks and the number of threads per thread-block,
following the same extended function-call syntax.
Figure~\ref{fig:MMMprogram} depicts the different types of components of 
an {\MMM} program.

\smallskip\noindent{\small \bf Scheduling and synchronization.}
At run time, an {\MMM} machine schedules thread-blocks 
(from the same or different kernels) onto SMs,
based on the dependencies specified by the edges of the DAG
and the hardware resources required by each thread-block.
Threads within a thread-block can 
cooperate with each other via the local memory of the
SM running the thread-block.
Meanwhile, thread-blocks interact with each other via the global memory.
In addition, threads within a thread-block 
are executed physically in parallel by an SM.
Moreover, the programmer cannot make any assumptions
on the order in which thread-blocks of a given
kernel are mapped to the SMs.
Hence, {\MMM} programs run correctly
on any fixed number of SMs.

\smallskip\noindent{\small \bf Memory access policy.}
All threads of a given thread-block can access 
simultaneously any memory cell of the local memory or the global memory:
read/write conflicts are handled by the 
CREW (concurrent read, exclusive write) policy.
However, read/write requests to the global memory
by two different thread-blocks cannot be executed
simultaneously. In case of simultaneous requests, 
one thread-block is chosen randomly and served first,
then the other is served.

For the purpose of analyzing program performance, 
we define two {\em machine parameters}:
\begin{itemize}
\item[$U$:] Time (expressed in  clock cycles) to transfer one machine word 
        between the global memory and the local memory of any SM.
\item[$Z$:] Size (expressed in machine words) of the local memory of 
        any SM. 
\end{itemize}
Thus, the quantity $1/U$ is a throughput measure and has the following
property. If $\alpha $ and $\beta$ are the numbers of words respectively 
read and written to the global memory by one thread of a thread-block $B$,
then the total time $T_D$ spent in data transfer between the global memory 
and the local memory of an SM executing $B$ satisfies
\begin{equation}
\label{eq:timefordata}
T_D \ \leq \ (\alpha + \beta) U.
\end{equation}
We observe that, on actual machines,
some hardware characteristics may
reduce data transfer time (like coalesced accesses to global memory)
or the contribution of the latter on the overall running time 
(like concurrent execution of thread-blocks on the same SM 
with fast context switching in order to hide data transfer latency).
Other hardware characteristics (like partition camping)  
may increase data transfer time.
However, this reduced  or increased transfer time will 
remain in  $O(\alpha + \beta)$.
Therefore, the fact that {\MMM} machines do not
have these hardware characteristics will not
lead us to incorrect asymptotic upper bounds 
when estimating the running time of {\MMM} programs.

Similarly, the local memory size $Z$ unifies 
different characteristics of an SM and, thus, of a thread-block.
Indeed, each of the following quantities is at most 
equal to $Z$:
the number of cores of an SM, the number of threads of a 
thread-block, the number of words in a data transfer
between the global memory and the local memory of an SM.

Relation~(\ref{eq:timefordata}) calls for another comment.
One could expect the introduction of a third machine parameter, say $V$, 
which would be the time to execute one {\em local operation}
(arithmetic operation, read/write in the local memory),
such that, if $\sigma$ is the total number of local operations
performed by one thread
of a thread-block $B$,
then the total time $T_A$ spent in {local operations}
by an SM executing $B$ would satisfy
\iflongversion
\begin{equation}
\label{eq:timeforcompute}
T_A \ \leq \ \sigma V.
\end{equation}
\else
$T_A \ \leq \ \sigma V.$
\fi
Therefore, for the total running time $T$ 
of the  thread-block $B$, we would have
\begin{equation}
\label{eq:threadtime}
T \ = \ T_A + T_D \ \leq \ \sigma V + (\alpha + \beta) U.
\end{equation}
Instead of introducing this third machine parameter $V$,
we let $V = 1$.
In other words, $U$ is the ratio between the time to transfer a
machine word and the time to execute a local operation.
We assume $U > 1$.

Finally, Relation (\ref{eq:threadtime}) requires another justification.
Indeed, this estimate suggests that, at every clock cycle, every thread
is either performing an arithmetic operation or accessing memory.
For this to be true, conditional branches responsible for
code divergence~\cite{Han:2011:RBD:1964179.1964184}
need to be eliminated
by techniques like code 
replication~\cite{Shin:2007:ICF:1299042.1299055},
such that all threads execute the same code path.
However, in a sake of clarity,
we will not perform this transformation
in the {\MMM} algorithms stated in the
subsequent sections.
In fact, we verified experimentally that  the impact of code divergence
on the performance of our algorithms is negligible.

\subsection{Many-core machine programs}
\label{sect:MMMprograms}

Recall that each {\MMM} program ${\cal P}$ is a 
DAG $({\cal K}, {\cal E})$,  
called the {\em kernel DAG} of ${\cal P}$,
where each node $K \in {\cal K}$ represents a kernel, 
and each edge $E \in {\cal E}$ records the fact that 
a kernel call  must precede another kernel call.
In other words, a kernel call can be executed 
provided that all its predecessors in the DAG $({\cal K}, {\cal E})$ 
have completed their execution.

\smallskip\noindent{\small \bf Synchronization costs.}
Recall that each kernel decomposes into 
thread-blocks and that all threads within a given kernel 
execute the same serial program, but with possibly different input data.
In addition, all threads within a thread-block 
are executed physically in parallel by an SM.
It follows that {\MMM} kernel code needs no synchronization statement,
like CUDA's {\tt \_\_syncthreads()}.
Consequently, the only form of synchronization taking place 
among the threads executing a given thread-block is that implied by code divergence.
This latter phenomenon can be seen as parallelism overhead.
As mentioned, an {\MMM} machine handles code divergence by 
eliminating the corresponding conditional branches via
code replication\footnote{
Conditional branch elimination in SIMD code is discussed
in~\url{https://www.kernel.org/pub/linux/kernel/people/geoff/cell/ps3-linux-docs/CellProgrammingTutorial/BasicsOfSIMDProgramming.html}.}
and the corresponding cost will be captured by the complexity
measures (work, span and parallelism overhead) of the {\MMM} model.
Since no synchronization occurs between thread blocks,
we turn our attention to scheduling.

\smallskip\noindent{\small \bf Scheduling costs.}
Since an {\MMM} abstract machine has infinitely many SMs
and since the kernel DAG defining an {\MMM} program ${\cal P}$ is
assumed to be known when ${\cal P}$  starts to execute,
scheduling ${\cal P}$'s kernels onto the SMs
can be done in time $O(\Gamma)$ where
$\Gamma$ is the total length of ${\cal P}$'s kernel code.
Thus, we shall neglect those costs in comparison
to the costs of transferring data between SMs' local memories
and the global memory.
We also note that assuming that, the kernel DAG 
is known when ${\cal P}$  starts to execute, allows us
to focus on parallelism overheads
resulting from this data transfer.
Extending {\MMM} machines to support programs
whose instruction stream DAGs unfold dynamically
at run time and integrating the resulting scheduling costs
as in~\cite{DBLP:conf/spaa/HeLL10} is left for future work.
\iflongversion
This {key observation} helps understanding the complexity
measures introduced in Section~\ref{sect:MMMmeasures}.
\fi

\smallskip\noindent{\small \bf Thread block DAG.}
Since each kernel of the program ${\cal P}$ decomposes into a finite
number of thread-blocks, we map ${\cal P}$ to
a second graph, called the {\em thread block DAG} of ${\cal P}$,
whose vertex set ${\cal B}({\cal P})$ consists of all thread-blocks of 
the kernels of ${\cal P}$ and such that $(B_1, B_2)$ is an edge 
if $B_1$ is a thread-block of a kernel preceding the kernel of 
the thread-block $B_2$ in ${\cal P}$.
This second graph defines two important quantities:
\begin{description}
\item[$N({\cal P})$:] number of vertices in the thread-block DAG of ${\cal P}$,
\item[$L({\cal P})$:] critical path length  
                     (where length of  a path is the number of edges in that path)
                      in the thread-block DAG of ${\cal P}$.
\end{description}

\iflongversion
For the purpose of analyzing program performance, 
we define five {\em program parameters}, 
summarized in Table~\ref{tab:programParameters}.

We also define five algorithm parameters, shown in table 2: $n$ is the input size of a many-core machine program; $ z $ is the maximum amount of the local memory per thread-block; $ q $ is the number of threads per thread-block; $ d $ is the number of thread-blocks needed during a parallel step; and $ {\pi} $ is the number of parallel steps of a many-core machine program. 

  \begin{table}[htdp]
  \begin{center} 
  \begin{tabular}{|c|l|}
  \hline
  Parameter & Description \\
  \hline
  $ n $ & Input size in machine words \\
  $ z $ & Maximum number of words of local memory \\
  & allocated per thread-block \\
  $ q $ & The number of threads per thread-block \\
  $ d $ & The maximum number of thread-blocks \\
  & in a parallel step \\
  $ {\pi} $ & The number of parallel steps \\
  \hline
  \end{tabular} 
  \caption{Algorithm parameters}
  \label{tab:programParameters}
  \end{center}
  \end{table}
\fi

\subsection{Complexity measures for the many-core machine model}
\label{sect:MMMmeasures}

Consider an {\MMM} program ${\cal P}$
given by its kernel DAG $({\cal K}, {\cal E})$.
Let $K \in {\cal K}$ be any kernel of ${\cal P}$
and $B$ be any thread-block of $K$.
We define the {\em work} of $B$, denoted by $W(B)$, as the total number
of local operations performed by all threads of $B$.
We define the {\em span} of $B$, denoted by $S(B)$, as the maximum
number of local operations performed by a thread of $B$.
Let $\alpha$ and $\beta$ be the maximum numbers of words 
read and written (from the global memory)
by a thread of $B$.
Then, we define the {\em overhead} of $B$, denoted by $O(B)$,
as $(\alpha + \beta)\,U$.
Next, the {\em work} (resp. {\em overhead}) $W(K)$ (resp. $O(K)$)
 of the kernel $K$ is 
the sum of the works (resp. overheads) of its thread-blocks,
while the {\em span} $S(K)$
of the kernel $K$ is the maximum
of the spans of its thread-blocks.

We consider now the entire program ${\cal P}$.
The {\em work} $W({\cal P})$ of ${\cal P}$ is defined as
the total work of all its kernels
\begin{equation*}
W({\cal P}) = \sum_{ K \in {\cal K} } \, W(K).
\end{equation*}
Regarding the graph $(K, E)$ as a weighted-vertex graph 
where the weight of a vertex $K \in {\cal K}$ is its span $S(K)$, 
we define the weight $S({\gamma})$ of any path ${\gamma}$ from the 
first executing kernel to the last executing kernel as 
$
S({\gamma}) = \sum_{K \in {\gamma} } \, S(K). 
$
Then, we define the {\em span} $S({\cal P})$ of the program ${\cal P}$ 
as 
\begin{equation*}
S({\cal P}) = \max_{{\gamma}} \, S({\gamma}).
\end{equation*}
\iflongversion
Regarding the graph $(K, E)$  as a weighted-vertex graph,
where the weight of a vertex $K$ is its 
 {\em overhead} $O(K)$, 
we define the {\em overhead} $O({\alpha})$ of 
an anti-chain ${\alpha}$ of $(K, E)$ as
\begin{equation*}
O({\alpha}) = \sum_{ K \in {\alpha} } \, O(K),
\end{equation*}
Finally, we define the {\em overhead} $O({\cal P})$ of ${\cal P}$ 
as the sum of the $O({\alpha})$'s  among all 
anti-chains ${\alpha}$  in $(K, E)$, that is,
\begin{equation*}
O({\cal P}) = \sum_{ {\alpha}}  \, O({\alpha}).
\end{equation*}
\fi
Finally, we define the {\em overhead} $O({\cal P})$ of the program ${\cal P}$
as the total overhead of all its kernels
\begin{equation*}
O({\cal P}) = \sum_{ K \in {\cal K} } \, O(K).
\end{equation*}
Observe that, according to Mirsky's theorem~\cite{M1971}, 
the number ${\pi}$ of parallel steps in ${\cal P}$ 
(i.e. anti-chains in $({\cal K}, {\cal E})$)
is equal to the maximum length of a path in $({\cal K}, {\cal E})$ from
the first executing kernel to the last executing kernel.

\subsection{A Graham-Brent theorem with  overhead}
\label{sect:GrahamBrent}

\begin{Theorem}
\label{thrm:GrahamBrendt}
We have the following estimate for the running time $T_{ P }$
of the program ${\cal P}$ when executed on {\P} SMs:
\begin{equation}
\label{eq:GrahamBrendt}
T_{\P} \leq (N({\cal P}) / {\P} + L({\cal P})) C({\cal P})
\end{equation}
where $C({\cal P}) = \max_{B \in {\cal B}({\cal P})} \, (S(B) + O(B))$,
\end{Theorem} 
The proof is similar to that of the original result.
One observes that the total number of {\em complete steps}
(for which $P$ thread-blocks can be scheduled by a greedy scheduler)
is at most $N({\cal P}) / {\P}$ while the
number of {\em incomplete steps} is at most $L({\cal P})$.
Finally, $C({\cal P})$ is an obvious upper bound for the 
running time of every step, complete or incomplete.

The proof of the following corollary follows from Theorem~\ref{thrm:GrahamBrendt}
and from the fact that costs of scheduling thread-blocks onto SMs are neglected.

\begin{Corollary}
\label{coro:Ning}
Let {\K} be the maximum number of thread blocks along an anti-chain
of the thread-block DAG of ${\cal P}$. Then the running time $T_{\cal P}$ 
of the program ${\cal P}$ satisfies:
\begin{equation}
T_{\cal P} \leq (N({\cal P}) / {\K} + L({\cal P})) C({\cal P})
\end{equation}
\end{Corollary}

Corollary~\ref{coro:Ning} allows us to estimate the running time
of an {\MMM} program as a function of the machine parameters
$Z$, $U$ and the thread-block DAG of ${\cal P}$. 
Thus this estimate does not depend on the number of SMs in use
to execute ${\cal P}$.

\iflongversion
\subsection{Justification of the many-core machine model}
\label{sect:MMMjustification}

In the modern GPU architectures, NVIDIA Fermi or Kepler,
it has a two-level, distributed thread scheduler.
At the chip level, a global work distribution engine schedules thread-blocks to various SMs, 
while at the SM level, each warp scheduler distributes warps of 32 threads to its execution units.
In this scenario, it supports fast context switching between threads and threads from different thread-blocks, 
or even between thread-blocks and thread-blocks among different kernels.
Besides, an amount of data required by a thread-block can be coalesced to 
transfer between global memory and local memory together.
Thus, we consider the overhead per thread-block as that caused by a thread,
and the machine parameter $U$ includes the time for fast context switching, 
coalesced memory access and latency.

On the other hand, regard to the new programming model of CUDA 5.0, 
a kernel that is ready can be launched by its predecessor at any time, 
while its predecessor waits to synchronize until the kernel has completed. 
Because of this feature, we can adjust it into a kernel DAG that our model deals with. 
Whenever a child kernel is called within its parent, we divide the parent kernel into
part $A$ and $B$, such that part $B$ of the parent kernel starts the same time as 
its child kernel, and part $A$ remains as the parent kernel itself.

  \begin{figure}[htb]
    \begin{center}
    \includegraphics[scale=0.4]{situation.eps}
    \end{center}
    \caption{Adjust any program into the DAG of many-core machine model}
    \label{fig:MMMadjust}
  \end{figure}
\fi

\iflongversion

In Section~\ref{sect:naivedivision}, we present a simple 
multithreaded algorithm computing $(q,r)$ on an {\MMM} machine.
We call this algorithm {\em naive} since it is a direct 
implementation of an idea that, at each division step,
each thread computes a coefficient of the next intermediate remainder.
In Section~\ref{sect:optimizeddivision}, we propose a second
{\MMM} algorithm with a goal of minimizing overhead.
We analyze both algorithms with the complexity measures
defined in Section~\ref{sect:MMMmeasures}.
In Section~\ref{sect:comparisondivision},
we compare these two algorithms  by means 
of Theorem~\ref{thrm:GrahamBrendt}.
In Section~\ref{cudaDivExpNTL},
we compare our  implementation of 
optimized plain division algorithm using CUDA with
division function found in NTL.

\else
\section{Polynomial division}
\label{sect:division}

Our first application of the {\MMM} deals 
with univariate polynomial division.
One division step, like a Gaussian elimination step,
performs a linear combination of two vectors.
Hence, the ideas developed in this section
would apply to linear algebra
and are not specific to polynomial arithmetic.

Let ${\KK}$ be a field and $a,b \in {\KK}[X]$
be univariate polynomials with coefficients in ${\KK}$, with $b \neq 0$.
Assume that  each arithmetic operation
(addition, subtraction, multiplication and division) in ${\KK}$
can be done with a single machine word operation on an {\MMM} machine.
Let $n$ and $m$ be non-negative integers such that 
we have ${\deg}(a) = n - 1$ and ${\deg}(b) = m - 1$.
Thus, $n$ and $m$ are the number of terms (null or not) 
of $a$ and $b$, respectively.
Let $q$ and $r$ be the quotient and the remainder in the
Euclidean division of $a$ by $b$.
Thus, $(q,r)$ is a unique couple of univariate polynomials
over ${\KK}$ such that $a = q \cdot b + r$ and ${\deg}(r) < {\deg}(b)$
both hold.

In Section~\ref{sect:naivedivision}, we present a 
multithreaded algorithm computing $(q,r)$ on an {\MMM} machine.
We call this algorithm {\em naive} since it implements  
the natural idea that, at each division step,
each thread computes one coefficient of the next intermediate remainder.
In Section~\ref{sect:optimizeddivision}, we propose a second
{\MMM} algorithm with the goal of minimizing data transfer.
We analyze both algorithms with the complexity measures
\footnote{See the detailed analysis in the form of
executable {\sc Maple} worksheet: 
\url{http://www.csd.uwo.ca/~nxie6/projects/mmm/division_overall.mw}}
of Section~\ref{sect:MMMmeasures}.
In Section~\ref{sect:comparisondivision},
we compare their running time estimates given by
Corollary~\ref{coro:Ning}.
\fi

\subsection{Naive algorithm}
\label{sect:naivedivision}

In each kernel call, each thread computes 
one coefficient of an intermediate remainder polynomial 
by means of one multiplication and one subtraction in the 
coefficient field ${\KK}$. 
Algorithm~\ref{alg:naivedivision} repeatedly
calls the kernel stated in Algorithm~\ref{alg:naivedivisionkernel}.
The latter performs one division step in parallel.
Let ${\ell}$ be the number of threads in a thread-block,
we note that each kernel uses $\lceil \frac{m}{\ell} \rceil$ thread-blocks.
We observe that each thread of a kernel 
reads/writes 3 to 5 words\footnote{
Indeed, in each thread-block, all active threads read
2 coefficients of $a$ and writes one back; 
moreover, the thread with ID $0$ computes 
and writes back one coefficient of $q$.}
in the global memory without storing them in the local memory.

  \begin{figure}[htb]
	\centering
  \includegraphics[scale=0.25]{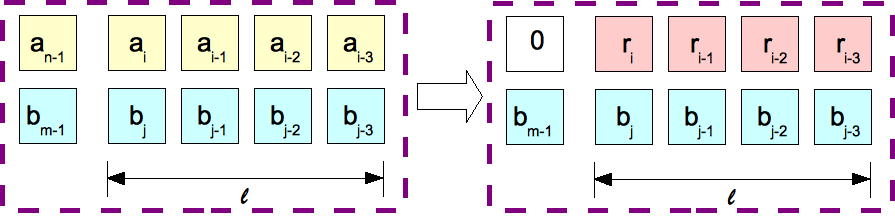}
  \caption{Naive division: illustration of a thread-block 
  reading coefficients from $a$,$b$ 
  and writing to $r$.}
  \label{fig:divnai}
  \end{figure}

We also notice that Algorithm~\ref{alg:naivedivision} performs exactly 
$n - m + 1$  consecutive calls to Algorithm~\ref{alg:naivedivisionkernel}.
Nevertheless, Algorithm~\ref{alg:naivedivisionkernel} works correctly
even if, after one division step,
the degree of an intermediate remainder drops by more than one.
This implementation choice is relevant to dense polynomials,
which are our primary interest.
In the sparse case, the degree of an intermediate remainder
needs to be computed after each division step.
Figure~\ref{fig:divnai} shows a division step within a thread-block
after running Algorithm~\ref{alg:naivedivisionkernel} once.



\begin{algorithm}[htb]
\caption{{\sf NaivePlainDivisionGPU($a,b$)}}
\label{alg:naivedivision}

\Indm
\KwIn{  $a, b \in {\KK}[X]$ with ${\deg}(a) \geq {\deg}(b)$ that is, $n-1 \geq m-1$.}
\KwOut{ $q, r \in {\KK}[X]$ s. t. $a = q \cdot b + r$
        and $\deg(r) < m-1$.}
\Indp
Let $\ell$ be the number of threads in a thread block\;    

Let $q$ be array of size $n-m+1$ with coefficients in {\KK}\; 


\For{$(i= (n-1) \ldots (m-1))$}
{      {\sf NaiveDivKernel}$\lll \lceil  m / {\ell} \rceil, \ell \ggg(a, b, q, i, m-1)$\; 
}

\If{ $a[0] == \cdots == a[m-1]  ==0$ }
{
{\bf return} $[q,0]$\;
}

Compute $k$, the maximum $i$ such that $a[i] \neq 0$ holds\;

Let $r$ be array of size $k+1$ s.t. $r[i] = a[i]$ for $0 \leq i \leq k$\;

{\bf return} $[q,r]$\;

\end{algorithm}

\begin{algorithm}[htb]
\caption{{\sf NaiveDivKernel($a,b, q, i, d_b$)}}
\label{alg:naivedivisionkernel}

\Indm
\KwIn{$a, b, q \in {\KK}[X]$,  $d_b=\deg(b)$, $i \in {\N}$, $d_b \leq i$.}
\Indp
Let {\sf blockID, blockDim, threadID} be the block id, number of threads per block, thread id respectively\;
$j = ${\sf blockID}$\cdot${\sf blockDim} $ + $ {\sf threadID}\;

\If{ $j \leq d_b$ }
{	

{
\If{ $j == 0$  }
{	
  \tcc{writing to global memory}
	$q[i-d_b] = (b[d_b])^{-1} \cdot a[i]$\;
	
}
}

\tcc{updating $a$ in global memory}
$a[j+i-d_b]$ -= $b[j] \cdot (b[d_b])^{-1} \cdot a[i]$\;

}
\end{algorithm}

We denote by $W_1$, $S_1$ and $O_1$,
the work, span and overhead of Algorithm~\ref{alg:naivedivision}, 
respectively.
Since each thread-block performs $2\,{\ell}+1$ arithmetic operations
and each thread makes at most 5 accesses to the global memory,
\iflongversion {
we obtain the following estimates, where ${\mu}$ stands for $n-m+1$,
\begin{equation*}
\label{eq:workspanoverheadnaidedivsion}
W_1 = \frac{{\mu}\,m\,(2\,{\ell}+1)}{\ell}, \,
S_1 = 3\,{\mu} \ \ {\rm and} \ \
O_1 = \frac{5\,{\mu}\,m\,U}{\ell}.
\end{equation*}
}
\else {
we obtain
$W_1 = \frac{(n-m+1)\,m\,(2\,{\ell}+1)}{\ell}$,
$S_1 = 3\,(n-m+1)$ and
$O_1 = \frac{5\,(n-m+1)\,m\,U}{\ell}$.
}
\fi
To apply Corollary~\ref{coro:Ning},
we shall compute the quantities $N({\cal P})$, $L({\cal P})$
and $C({\cal P})$ defined in Section~\ref{sect:MMM}.
We denote them here by $N_1$, $L_1$ and $C_1$, respectively.
One can easily check that we have
\iflongversion {
\begin{equation*}
\label{eq:brentnaidivision}
N_1 = \frac{{\mu}\,m}{\ell}, \,
L_1 = {\mu} \ \ {\rm and} \ \
C_1 = 3+5\,U.
\end{equation*}
}
\else {
$N_1 = \frac{(n-m+1)\,m}{\ell}$,
$L_1 = n-m+1$ and
$C_1 = 3+5\,U$.
}
\fi

\iflongversion

In Figure~\ref{fig:naiDiv}, we show how 
one naive division step is done on GPU 
by our algorithm.

  \begin{figure}[htb]
    \begin{center}
    \includegraphics[scale=0.6]{nai.eps}
    \end{center}
    \caption{A naive division step.}
    \label{fig:naiDiv}
  \end{figure}

\fi

\subsection{Optimized algorithm}
\label{sect:optimizeddivision}

\begin{figure}[htb]
	\centering
  \includegraphics[scale=0.25]{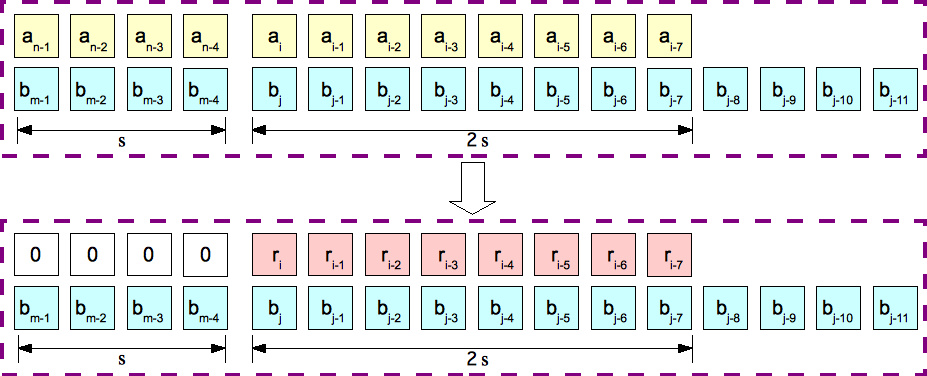}
  \caption{Optimized division: illustration of a thread-block 
  reading coefficients from $a$,$b$ 
  and writing to $r$.}
  \label{fig:divopt}
  \end{figure}

In each kernel, each thread updates a number of coefficients
(instead of just one) of an intermediate remainder polynomial 
repeatedly during a number of division steps,
thus without synchronizing data with other thread-blocks.
The motivation of this new scheme is to minimize
the amount of data transferred between global
and local memories.
Similarly to the scheme in Section~\ref{sect:naivedivision},
Algorithm~\ref{alg:Optimizedivision} repeatedly
calls Algorithm~\ref{alg:optdivisionkernel}.
Figure~\ref{fig:divopt} shows $s$ division steps within a thread-block
after running Algorithm~\ref{alg:optdivisionkernel} once.
More specifically, given an integer $s \geq 1$,
Algorithm~\ref{alg:optdivisionkernel} performs sufficiently 
many division steps (at most $s$) 
such that the output polynomial $r$
is either zero or its degree is less than that of $a$ at least by $s$.
To this end, each thread-block
\begin{itemizeshort}
\item uses $3\,s$ threads, 
\item loads the  coefficients of $X^d$, $X^{d-1}$, $\ldots$, $X^{d-s+1}$
      from $a$ (resp. $b$), that we call the $s$-{\em head} of $a$ (resp. $b$),
      where $d$ is the degree of $a$ (resp. $b$), see Lines 3-4, 
\item loads $2\,s$ (resp. $3\,s$) consecutive coefficients of $a$ (resp. $b$), say
       $X^{d_1}$, $X^{d_1-1}$, $\ldots$, $X^{d_1-2s+1}$
       ($X^{d_2}$, $X^{d_2-1}$, $\ldots$, $X^{d_2-3s+1}$) for some integer $d_1 > 0$
       (resp. $d_2 > 0$) which depends on the thread and thread-block IDs.
\end{itemizeshort}

The $s$-heads of $a$ and $b$ are used to keep track of the
leading coefficient of an intermediate remainder during
the entire execution of a kernel, see Lines 14-15
of Algorithm~\ref{alg:optdivisionkernel}.
This task is achieved by the first $s$ threads of a thread-block
and requires $s + (s-1) + \cdots + 1 = \frac{s(s+1)}{2}$ arithmetic
operations.
Meanwhile, the other $2\,s$ threads update $2 s$ coefficients
of $a$, see Lines 16-17, which takes $2 \, s \cdot 2 \, s$
arithmetic operations.
Finally, at Lines 12-13 (resp. 18-19) the quotient $q$ (resp. the
intermediate remainder $a$) is updated in the global memory.

We denote the work, span and overhead of the optimized algorithm
by $W_s$, $S_s$ and $O_s$, respectively.
Since each thread makes at most 9 accesses to the global memory,
\iflongversion {
we obtain the following estimates, where ${\mu}$ stands for $n-m+1$,
\begin{equation}
\label{eq:workspanoverheadoptimizeddivsion}
W_{\rm opt} = \frac{{\mu}\,m\,(9\,s+1)}{4\,s}, \,
S_{\rm opt} = 3\,{\mu} \ \ {\rm and} \ \
O_{\rm opt} = \frac{9\,{\mu}\,m\,U}{2\,s^2}.
\end{equation}
}
\else {
we obtain 
$W_s = \frac{(n-m+1)\,m\,(9\,s+1)}{4\,s}$,
$S_s = 3\,(n-m+1)$ and
$O_s = \frac{9\,(n-m+1)\,m\,U}{2\,s^2}$.
}
\fi
To apply Corollary~\ref{coro:Ning},
we shall compute the quantities $N({\cal P})$, $L({\cal P})$
and $C({\cal P})$ defined in Section~\ref{sect:MMM}.
We denote them here by $N_s$, $L_s$ and $C_s$, respectively.
One can easily check that we have
\iflongversion {
\begin{equation}
\label{eq:brentoptdivision}
N_s = \frac{{\mu}\,m}{2\,s^2}, \,
L_s = \frac{{\mu}}{s} \ \ {\rm and} \ \
C_s = 3\,s+9\,U.
\end{equation}
}
\else{
$N_s = \frac{(n-m+1)\,m}{2\,s^2}$,
$L_s = \frac{n-m+1}{s}$ and
$C_s = 3\,s+9\,U$.
}
\fi

\iflongversion

In Figure~\ref{fig:optDiv}, we show how 
optimized division steps is done on GPU 
by our algorithm.

  \begin{figure}[htb]
    \begin{center}
    \includegraphics[scale=0.6]{opt.eps}
    \end{center}
    \caption{Optimize division steps.}
    \label{fig:optDiv}
  \end{figure}

\fi

\begin{algorithm}[htb]
\caption{{\sf OptimizePlainDivisionGPU($a,b, s$)}}
\label{alg:Optimizedivision}
\Indm
\KwIn{  $a, b \in {\KK}[X]$ with ${\deg}(a) \geq {\deg}(b)$ that is, $n-1 \geq m-1$ and  $s \in {\N}$.
}

\KwOut{ $q, r \in {\KK}[X]$ s. t. $a = q \cdot b + r$
        and $\deg(r) < m-1$.}


\Indp
Let $\ell= 3s$ be the number of threads in a thread block\;
Let $q$ be array of size $n-m+1$ with coefficients in {\KK}\; 

\For{$(i= n-1; i \geq m-1; i = i - s)$}
 {    \sf OptDivKer$\lll \lceil  m/(2s) \rceil, \ell \ggg (a, b, q, i, m-1, s$)\; 
}

\If{ $a[0] ==\cdots == a[m-1]  ==0$ }
{
{\bf return} $[q,0]$\;
}

Compute $k$, the maximum $i$ such that $a[i] \neq 0$ holds\;

Let $r$ be array of size $k+1$ s.t. $r[i] = a[i]$ for $0 \leq i \leq k$\;

{\bf return} $[q,r]$\;
\end{algorithm}

\begin{algorithm}[htb]
\caption{{\sf OptDivKer$(a,b, q, i, d_b, s)$}}
\label{alg:optdivisionkernel}

\Indm
\KwIn{$a, b, q \in {\KK}[X]$,  $i  \in {\N}$,  $d_b=\deg(b)$ and $s \in {\N}$.}


\Indp


Let {\sf sAc,  sBc, sA, sB}  be {\sf local } arrays of size $s,s,2s,3s$ respectively each  with coefficients in {\KK}\;

$j = ${\sf blockID}$\cdot${\sf blockDim} $+$ {\sf threadID}; $t = $ {\sf threadID}\;

\tcc{Reading from global memory}
\If{$t <s$}
{	 {\sf sAc}$[t]= a[i-t]$;{ } {\sf sBc}$[t]= b[d_b -t]$;{ }	{\sf sB}$[t]    = b[d_b - 2s\,${\sf blockID}$ - t]$\;
}

\If{$t\geq s $}
{	 {\sf sA}$[t-s] = a[ i  - s - 2s\,${\sf blockID}$ - t]$;{ }	 {\sf sB}$[t] = b[d_b - 2s\,${\sf blockID}$ - t]$\;
}



\For{$(k= 0; (k < s) \wedge (i+k \geq d); k = k+1)  $}
{	\While { $(k < s) \wedge ($ {\sf sAc}$[k] == 0)$}
	{	 $k = k+1$\;
	}
	\If {  $k \geq s$}
	{	 {\bf break}\;
	}

	\If{$j == 0$}
	{
	  \tcc{Writing $q$ to global memory}	
		$q[i - d_b - k]$ = {\sf sAc}$[k] \cdot b[d_b]^{-1}$\;
	}
	\If{$k \leq t < s $}
	{	
 		{\sf sAc}$[t]$ -= {\sf sBc}$[t-k]$ $\cdot$ {\sf sAc}$[k] \cdot b[d_b]^{-1}$\;
	}
	\If{$ t \geq s$}
	{	 {\sf sA}$[t-s]$ -= {\sf sB}$[t-k]$ $\cdot$ {\sf sAc}$[k] \cdot b[d_b]^{-1}$\;
	}
}

\If{$t\geq s $}
{	
  \tcc{Writing  back $a$ to global memory}
  $a[ i  - s - 2s\,${\sf blockID}$ - t] = $ {\sf sA}$[t-s]$\; 
}

\end{algorithm}

\subsection{Comparison of running time estimates}
\label{sect:comparisondivision}

Before following the algorithm optimization strategy stated in the introduction,
we replace ${\ell}$ and $s$ by $Z/2$ and $Z/7$, respectively, since
$2 {\ell}$ or $7 s$ coefficients must fit into the local memory, 
that is, $2 {\ell} \leq Z$ and $7 s \leq Z$.
Now, we observe that the work ratio $W_1 / W_s$ 
is asymptotically constant:
\begin{equation}
\frac{W_1}{W_s} = \frac{8\,(Z + 1)}{ 9\,Z + 7 },
\end{equation}
and $S_1 / S_s = 1$.
Then, we compute the overhead ratio:
\begin{equation}
\frac{O_1}{O_s} = {\frac {20}{441}}\,Z.
\end{equation}
We observe that this substantial improvement in data transfer 
overhead  is done at 
a fairly low expense in terms of work overhead.
Next, applying Corollary~\ref{coro:Ning},
the running time estimate of the naive and optimized algorithms
are bounded over, respectively by
\iflongversion

\begin{equation}
\label{eq:upperboundsdivision1}
(N_1 / {\K} + L_1) \cdot C_1 \ \ {\rm and} \ \ 
(N_s / {\K} + L_s) \cdot C_s.
\end{equation}
\else
$T_1 = (N_1 / {\K_1} + L_1) \cdot C_1$ with $\K_1 = \frac{m}{\ell}$ and
$T_s = (N_s / {\K_s} + L_s) \cdot C_s$ with $\K_s = \frac{m}{2\,s}$.
\fi
\iflongversion
\begin{equation}
\label{eq:upperboundsdivisionratio}
\frac{2}{3}\,{\frac { \left( 3+5\,U \right)  \left( m+{\ell}\,P \right) {s}^{2}}{{\ell}\,
 \left( s+3\,U \right)  \left( m+2\,s\,P \right) }}.
\end{equation}
\begin{equation*}
\label{eq:upperboundsdivisionratio2}
\frac{2}{3}\,{\frac { \left( 3+5\,U \right)  \left( 2\,m+Z\,P \right) Z}{
 \left( Z+21\,U \right)  \left( 7\,m+2\,Z\,P \right) }}.
\end{equation*}

$\frac{2}{3}\,{\frac { \left( 3+5\,U \right)  \left( 2\,m+Z\,P \right) Z}{
 \left( Z+21\,U \right)  \left( 7\,m+2\,Z\,P \right) }}.$
\fi
We compute the ratio $R = T_1 / T_s$, that is,
\begin{equation}
\label{eq:upperboundsdivisionratio3}
R = \frac{(3+5\,U)\,Z}{3\,(Z+21\,U)}.
\end{equation}
We observe that $R$ is larger than $1$ if and only
if $Z > 12.6$ holds.
The above condition clearly holds on actual GPU architectures.
Thus the optimized algorithm (that is for $s > 1$) 
is overall better than the naive one (that is for $s = 1$).
This is verified in practice in~\cite{HM2012}, where $s$ is set to $512$.

\section{Radix sort}
\label{sect:radix}

In~\cite{SHG2009}, the authors present a CUDA implementation of the
radix sort algorithm.
Assuming that all entries are non-negative integers of bit-size $c$,
this CUDA implementation sorts $n$ entries by performing $\frac{c}{s}$  {\em passes}
where  $s$ is a program parameter.
In each pass, each thread-block first loads and sorts its tile using $s$ iterations 
of 1-bit split,
and write back its $2^s$-entry digit histogram and sorted data.
Then, it performs a prefix sum over the histogram stored in a column-major order.
Finally, each thread-block copies its elements to the correct output position.

Let ${\ell}$ be the number of threads per thread-block.
Following~\cite{SHG2009}, we assume that each thread
deals with $4$ elements.
Then, for each thread-block, $4\,{\ell}$ original elements, $4\,{\ell}$ sorted elements 
and a $2^s$-entry digit histogram 
must fit into the local memory, hence  $8\,{\ell}+2^s \leq Z$.
Thus, the maximum overhead per thread-block is $9\, U$ (loading $4$ elements, 
writing back $4$ sorted elements and $1$ value of the histogram
\footnote{See the detailed analysis in the form of
executable {\sc Maple} worksheet: 
\url{http://www.csd.uwo.ca/~nxie6/projects/mmm/sorting_overall.mw}}).

We compute the work, span and overhead, respectively as 
$W_{s} = c \left( \frac {22\,s\,{\ell}+s+12}{4\,s\,{\ell}} + \frac {{2}^{s}+20\,{2}^{s}\,{\ell}}{16\,s\,{\ell}^{2}} + 1 \right) n + \frac {c \, ( 16+192\,{\ell} ) }{16\,s}$,
$S_{s} = c \left(8\,\log_2 {\ell} + \frac{16}{s} \log_2 {\ell} + 41 + \frac{54}{s}\right)$, 
$O_{s} = c\,U \left( \frac {9}{2\,s\,{\ell}}\,n +\frac{17\,2^s}{16\,s\,{\ell}^2}\,n-\frac{1}{s} \right)$.

We view the case $s=1$ as a naive radix sort algorithm, with  work, span and overhead  
given by $W_1$, $S_1$ and $O_1$, respectively. 
Letting $n$ escaping to infinity, the work ratio ${W_{1}}/{W_{s}}$ 
is asymptotically equivalent to:
\begin{equation}
\frac{W_{1}}{W_{s}} \ \sim \ \frac{104\,s\,{\ell}^2 + 92\,s\,{\ell} + 2\,s}{88\,s\,{\ell}^2+16\,{\ell}^2+20\,2^s\,{\ell}+4\,s\,{\ell}+48\,{\ell}+2^s}.
\end{equation}
Similarly, the span ratio 
$S_{1}/S_{s} = \frac{s\,(24\,\log_2{\ell}+95)}{(8\,s+16)\,log_2{\ell}+41\,s+54}$
is asymptotically constant,
meanwhile the overhead ratio is:
\begin{equation}
\frac{O_{1}}{O_{s}} \ \sim \ \frac{s\,(72\,{\ell}+34)}{72\,{\ell}+17\,2^s}.
\end{equation}
We notice that if $2^s = \Theta({\ell})$ holds, 
we can reduce the overhead by a factor of $s$ 
while increasing the work by a constant factor only. 
Meanwhile, with $2^s=\Theta({\ell}^2)$, 
we increase both the work and the overhead by a non-constant factor.
In this latter scenario, we could not
optimize the naive algorithm in any sense.

To apply Corollary~\ref{coro:Ning},
we compute the three quantities 
$N_{s} = \frac {c}{s} \left({\frac {1}{2\,{\ell}}}+{\frac {{2}^{s}}{8\,{\ell}^{2}}} \right)\,n$,
$L_{s} = \frac{5\,c}{s}$ and
$C_{s} = s \left( 41 + \right.$ $\left. 8\, \log_2 {\ell} \right) + 12 +9\,U$.
\iflongversion
We also denote these three quantities $N_1$, $L_1$ and $C_1$ 
for the naive algorithm, that is, when $s=1$.
\fi
Then, the  ratio $R$ of the running time estimate
between the naive algorithm and that for an arbitrary $s$ is:
\begin{equation*}
R = \frac{(14\,{\ell}+2)\,(53+8\,\log_2{\ell}+9\,U)\,s}
{(14\,{\ell}+2^s)\,(41\,s+8\,s\,\log_2{\ell}+12+9\,U)}.
\end{equation*}
We then replace $s$ by $\log_2 {\ell}$,
since we would like to determine
whether the overall running time is better or worse in this case.
The  quotient of the leading terms in $R$ becomes
\begin{equation}
\frac{7\,{\ell}\,\log_2{\ell}\,(8\,\log_2{\ell}+9\,U)}
{60\,{\ell}\,\log_2^2{\ell}}.
\end{equation}
This  ratio is larger than $1$ 
for ${\ell} < 2^{15.75\,U}$, which is realistic.
Therefore, letting $2^s = \Theta({\ell})$,
the data transfer overhead is reduced by a factor of $s$ 
and leads to an optimized algorithm.
This is consistent 
with  the empirical results of~\cite{SHG2009}.

\section{Polynomial multiplication}
\label{sec:mul}

\iflongversion
Let ${\KK}$ be a field and $a,b \in {\KK}[X]$
be two univariate polynomials over ${\KK}$
and with variable $X$ (like in Section~\ref{sect:division}).
Let $n$ and $m$ be positive integers such that 
${\deg}(a) = n - 1$ and ${\deg}(b) = m - 1$.
\else
Let $a$ and $b$ be polynomials as in Section~\ref{sect:division},
\fi
and $f = a \times b$ be their product.
Our multiplication algorithm is based on the well-known {\em long
multiplication}\footnote{\url{http://en.wikipedia.org/wiki/Multiplication_algorithm}}
and consists of two phases.
During the {\em multiplication phase}, every coefficient of $a$ 
is multiplied with every coefficient of $b$;
the resulting products are accumulated in an intermediate array, denoted by $M$.
Then, during the {\em addition phase}, these accumulated products
are added together to form the polynomial $f$.

For this application, 
the program parameter (as defined
in the introduction) is an integer $s > 0$, 
representing for each thread-block, the number of coefficients of $b$ 
to be multiplied by a number of coefficients of $a$ in the coefficients
multiplication phase, as well as the number of sums per thread in
the addition phase.
\footnote{See the detailed analysis in the form of
executable {\sc Maple} worksheet: 
\url{http://www.csd.uwo.ca/~nxie6/projects/mmm/multiplication_overall.mw}}
Algorithm~\ref{algo:PlainMulGPU} is the top level algorithm:
it performs the multiplication phase via Algorithm~\ref{algo:mulKernel},
and the addition phase by repeated calls to 
Algorithm~\ref{algo:MulAdd}.

\begin{figure}[htb]
	\centering
  \includegraphics[scale=0.25]{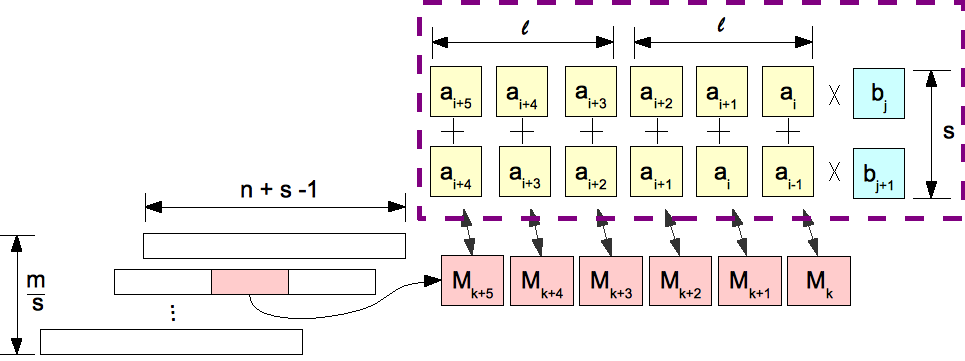}
  \caption{Multiplication phase: illustration of a thread-block 
  reading coefficients from $a$,$b$ and writing to
  the auxiliary array $M$.}
  \label{fig:mulmul}
  \end{figure}

We denote by ${\ell}$ the number of threads per thread-block.
In multiplication phase, 
each thread-block reads  $s\,{\ell} + s - 1$ coefficients of $a$,
$s$ coefficients of $b$, computes ${\ell}\,s^2$ products,
followed by ${\ell}\,s\,(s-1)$ of additions. Thus, each thread-block contributes
$s\,{\ell}$ {\em partial sums} to the two-dimensional array $M$, 
whose format is $x \cdot y$, where $x = \frac{m}{s}$ and $y=n+s-1$.
This multiplication phase, illustrated by Figure~\ref{fig:mulmul},
loads $s\,{\ell} +s -1$ coefficients of $a$ to guarantee the correctness
of the results in $M$.

\begin{figure}[htb]
	\centering
  \includegraphics[scale=0.25]{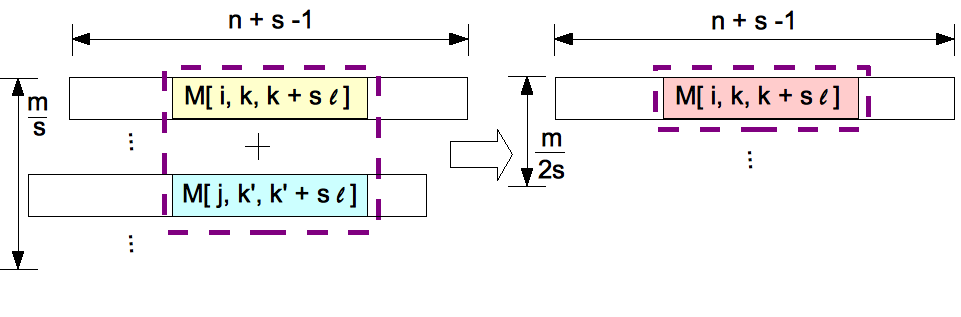}
  \caption{Addition phase: illustration of a thread-block
  reading and writing to the auxiliary array $M$.}
  \label{fig:muladd}
  \end{figure}

In the addition phase, the $x$ rows of the auxiliary array $M$
are added pairwise in $\log_2{x}$ parallel steps. After each step, the 
number of rows in $M$ is reduced by half, until we obtain only one row
that is, $f = a \times b$. To be more specific,
when adding rows $i$ and $j$ (for $i < j$) at a given parallel step,
shown in Figure~\ref{fig:muladd}, 
each thread-block loads 
$s\,{\ell}$ elements of $M[i]$ and $M[j]$, respectively, and then 
adds $M[j]$ to $M[i]$.

\begin{algorithm}[htb]{}
\caption{{\sf PlainMultiplicationGPU}($a,b,f, s$) }
\label{algo:PlainMulGPU}

\Indm
\KwIn{$a, b \in {\KK}[X]$ with $n-1:={\deg}(a)$ and $m-1:= {deg}(b)$ and an integer $s \geq 1$.}

\KwOut{$f \in {\KK}[X]$ and $f=a \times b$.}

\Indp
$y = n+s-1$;
 $x = m/s$\;

Let $M$ be an array of size $x \cdot y$ with coefficients in {\KK}\; 

$\ell$ is the number of threads per block\;

 {\sf MulKer}$\lll x \cdot y/(s \cdot {\ell}), \ell \ggg (a,b,M,n,m,s)$\;
\For{($i = 0; i \leq \log_2 {s}; i= i +1$)}
{
	 {\sf AddKer}$\lll x \cdot y/(2^{i+1}\,s \cdot {\ell}), \ell \ggg (M,f, y,s, x, i)$\;
}
 {\bf return} { }$f$\;
\end{algorithm}

\begin{algorithm}[htb]
\caption{{\sf MulKer}$(a,b,M,n,m,s)$}
\label{algo:mulKernel}

\Indm
\KwIn{ $a, b, M \in {\KK}[X]$ and an integer $s \geq 1$. }

\Indp
${\ell} = $ {\sf blockDim}; $t = $ {\sf threadID};
$j = ${\sf blockID}$\cdot {\ell} + t$\;
Let $B'$ and $A'$ be two {\sf local } arrays in {\KK} of size $s$ and 
${\ell} \cdot s + s-1$ respectively\;

\tcc{copying from global}
$i = s \cdot \lfloor s \cdot j / (n+s-1) \rfloor + t$\;
\If {$i< m \wedge t < s$} {
  $B'[t] = b[i]$\;
}
$i = s \cdot (j \mod \frac{n+s-1}{s})$\;
\For {($k = 0; k < s; k = k+1$)} {
  \If {$i+k\cdot{\ell}+t < n$} {
    $A'[k\cdot{\ell}+t] = a[i+k\cdot{\ell}+t]$\;
  }
}
\If {$i - s+t >0 \wedge t < s-1$} {
  $A'[{\ell}\cdot s+t] = a[i - s+t]$\;
}
\ElseIf {$t < s-1$} {
  $A'[{\ell}\cdot s+t] = 0$\;
}

\For {($e=0; e < s; e = e+1$)} {
  $h = 0$\;
  \For {($k = 0; k < s; k = k+1$)} {
    $h$ += $ A'[e \cdot {\ell}+k] \cdot B'[k] $\;
  }
  \tcc{writing to global memory}
  $M[s \cdot j + e] = h$\;
}

\end{algorithm}

\begin{algorithm}[htb]
\caption{{\sf AddKer}$(M,f, y, s, x,i)$ }
\label{algo:MulAdd}
\Indm
\KwIn{   $M, f,  \in {\KK}[X]$ and $y,s,x, i$ are positive integers.}
\Indp

$j = ${\sf blockID}$\cdot${\sf blockDim} + {\sf threadID}\;
$h = s \cdot j \mod y$\;
$k = 2^i-1+2^{i+1}\,\lfloor s \cdot j / y \rfloor$\;

\If {$h < 2^i\,s$} {
  \For {($e = 0; e < s; e = e+1$)} {
     $f[k\cdot s+h+e]$ += $M[k \cdot x+h+e]$\;
  }
}
\Else {
  \For {($e = 0; e < s; e = e+1$)} {
    $M[(k+ 2^i) \cdot x+h-2^i \, s+e]$ += $M[k \cdot x+h+e]$\;
	}
}

\end{algorithm}

\iflongversion
We denote by $W_{\rm s}$, $S_{\rm s}$ and, $O_{\rm s}$, the work, span and overhead, respectively. Each thread-block performs $t\,(2\,s-1)$ arithmetic operations in multiplication phase, and ${\ell}$ arithmetic operations in addition phase. Each thread makes at most $3$ accesses to the global memory. We obtain the following estimates, where ${\delta}$ stands for $n+s-1$ and ${\mu}$ stands for $2\,s-1$,
\begin{equation}
\label{eq:workspanoverheadxmul}
W_{s} = (2\,m-\frac{1}{2})\,{\delta},\,
S_{s} = {\mu}+\log_{2} \frac{m}{s}, \,
O_{s} = \frac{3\,{\delta}\,(2\,m-s)\,U}{s\,{\ell}}.
\end{equation}

In order to apply Theorem~\ref{thrm:GrahamBrendt},
we shall compute the quantities $N({\cal P})$, $L({\cal P})$
and $C({\cal P})$ defined in Section~\ref{sect:MMM}.
We denote them here by $N_{\rm s}$, $L_{\rm s}$ and $C_{\rm s}$, respectively.
One can easily check that we have
\begin{equation}
\label{eq:brentxmul}
N_{s} = \frac{{\delta}\,(2\,m-s)}{s\,{\ell}},\,
L_{s} = \log_{2} \frac{m}{s} + 1 \ \ {\rm and} \ \
C_{s} = {\mu}+3\,U.
\end{equation}
\else
Considering any thread-block of the multiplication phase, 
we notice that $s$ coefficients of $b$ and $s\,{\ell}+s-1$ 
coefficients of $a$ are loaded
and $s\,{\ell}$ results are written back to global memory.
Hence $2\,s\,{\ell}+2\,s-1$ coefficients 
must fit into local memory, that is, we have $2\,s\,{\ell}+2\,s-1 \leq Z$.
Next, we compute the work, span and parallelism overhead, respectively as
$W_{s} = (2\,m-\frac{1}{2})\,(n+s-1)$,
$S_{s} = 2\,s^2+s\,\log_{2} \frac{m}{s}-s$ and
$O_{s} = \frac{(n+s-1)\,(5\,m\,s+2\,m-3\,s^2)\,U}{s^2\,{\ell}}$.
We also obtain the quantities characterizing the thread block DAG
that are required in order to apply Corollary~\ref{coro:Ning}:
$N_{s} = \frac{(n+s-1)\,(2\,m-s)}{s^2\,{\ell}}$,
$L_{s} = \log_{2} \frac{m}{s} + 1$ and
$C_{s} = s\,(2\,s-1)+2\,U\,(s+1)$.
\fi

\iflongversion
{
For this application, we use ``naive algorithm'' as
the one obtained by setting $s=1$ in Algorithm~\ref{algo:PlainMulGPU}.
For fixed $n$ and $m$, we have the work ratio is asymptotically constant:
\begin{equation}
\frac{W_{\rm 1}}{W_{\rm s}} = {\frac {n}{n+s-1}},
\end{equation}
 and the overhead ratio increases as $s$ increases, that is,
\begin{equation}
\frac{O_{\rm 1} }{ O_{\rm s} } =
{\frac {n\, \left( 2\,m-1 \right) s}{ \left( n+s-1 \right)  \left( 2\,
m-s \right) }}.
\end{equation}
Thus, we observe that this substantial improvement is done at 
a fairly low expense in terms of work overhead.
Applying Corollary~\ref{coro:Ning},
the running times on $p$ 
SMs of the naive algorithm and the algorithm with arbitrary $x$ are 
bounded over by
\begin{equation}
\label{eq:upperboundsdivision2}
(N_{\rm 1} / P + L_{\rm 1}) \cdot C_{\rm 1} \ \ {\rm and} \ \ 
(N_{\rm s} / P + L_{\rm s}) \cdot C_{\rm s}.
\end{equation}
 and replace $m$ by $n$.
When $n$ escapes to infinity, the ratio $R$ is equivalent to
\begin{equation}
\label{eq:upperboundsmulratio}
\frac{(1+3\,U)\,x}{2\,x-1+3\,U}.
\end{equation}
One can assume $3\,U > 1$, which implies that the above
ratio is always greater than $1$ as soon as $ s > 1$ holds.
Therefore, the algorithm with arbitrary $s$ outperforms the naive one.
}
\else
{
We set $s=1$ in Algorithm~\ref{algo:PlainMulGPU}, and view it
as a ``naive algorithm".
The work ratio $W_{1} / W_{s} = {\frac {n}{n+s-1}}$, 
is asymptotically constant as $n$ escapes to infinity.
The span ratio 
$S_{1} / S_{s} = \frac{\log_2{m}+1}{s\,(\log_2{(m/s)}+2\,s-1)}$
shows that $S_s$ grows asymptotically with $s$.
The parallelism overhead ratio, letting $m = n$:
\begin{equation}
\frac{O_{1} }{ O_{s} } =
\frac{n\,s^2\,(7\,n-3)}{(n+s-1)\,(5\,n\,s+2\,n-3\,s^2)}.
\end{equation}
We observe that, as $n$ escape to infinity, this latter ratio
is asymptotically equivalent to $s$.
Applying Corollary~\ref{coro:Ning},
let $R$ be the ratio of the running time estimate
between the naive algorithm and that for an arbitrary $s$.
We obtain
\begin{equation}
\label{eq:upperboundsmulratio}
R = \frac{(n\,\log_2{n}+3\,n-1)\,(1+4\,U)}
{(n\,\log_2{\frac{n}{s}}+3\,n-s)\,(2\,U\,s+2\,U+2\,s^2-s)},
\end{equation}
which is essentially $\frac{2\log_2{n}}{s\,\log_2{(n/s)}}$.
This latter ratio is smaller than $1$, 
such that the ``initial" algorithm 
(that is for $s=1$) performs better. 
This also indicates that increasing $s$ makes the algorithm performance
worse.
In practice shown in Table~\ref{tab:mul}, setting $s=4$ 
performs best, while with larger $s$,
the running time becomes slower, which is coherent with our theoretical
analysis.

}
\fi

\section{The Euclidean algorithm \\ for polynomials}
\label{sect:gcd}

Let $a$ and $b$ be polynomials as defined in Section~\ref{sect:division}.
In Section~\ref{sect:naivegcd}, we present a simple multithreaded algorithm 
that computes {\sf GCD}$(a, b)$ on an {\MMM} machine.
We call it naive (like Algorithm~\ref{alg:naivedivision})
as this algorithm also performs one division step within one kernel.
In Section~\ref{sect:optgcd}, we describe another algorithm on an {\MMM} machine
that reduces the overhead of data transfer by performing several division steps
within one kernel. Finally, in Section~\ref{sect:analysisgcd}
we compare
 those two algorithms by 
means of Corollary~\ref{coro:Ning}.
A detailed analysis  in the form of executable {\sc Maple} worksheet
is available at\footnote{\url{http://www.csd.uwo.ca/~nxie6/projects/mmm/euclidean_overall.mw}}.

\subsection{Naive algorithm}
\label{sect:naivegcd}

\begin{figure}[htb]
	\centering
	\includegraphics[scale=0.25]{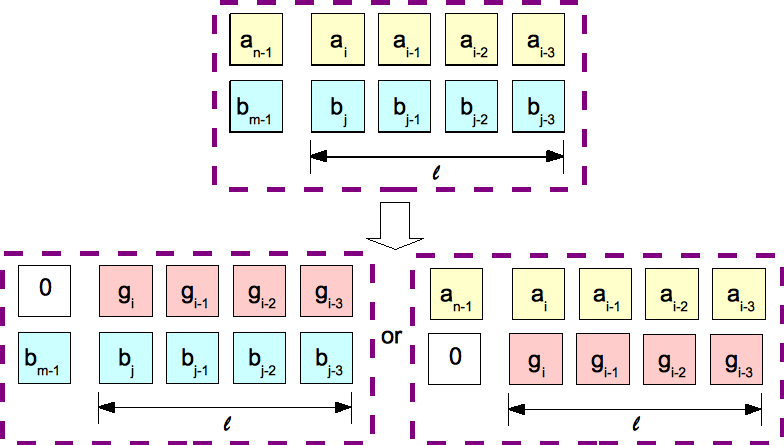}
  \caption{Naive Euclidean: illustration of a thread-block reading
  coefficients from $a$,$b$ and writing to $g$.}
  \label{fig:gcdnai}
  \end{figure}

Algorithm~\ref{alg:naivegcd} computes
{\sf GCD}$(a,b)$. Similarly to the naive
division of Algorithm~\ref{alg:naivedivision}, it calls a kernel
(stated as Algorithm~\ref{alg:naivGcdnkernel})  that completes one division step.
The former algorithm calls the latter one
at most $n+m-2$ times. 
Algorithm~\ref{alg:naivGcdnkernel} is the same as
Algorithm~\ref{alg:naivedivisionkernel} except that
it checks the current degrees of both $a$
and $b$ so as to decide which one
 takes the role of the divisor, and
then it completes a division step.
To do so, we use an array {\sf st} of length $2$ to store
the degrees of $a$ and $b$ during each division step.
\iflongversion
Each thread also reads/writes  
$3$ to $5$ words in the global memory without storing then in the local memory, and
it computes
one coefficient of an intermediate
remainder. So, the number of 
active threads in a division step
depends on the degree of the divisor 
polynomial. The active thread, whose
rank is maximum, updates {\sf st} array 
at the end of the division step correctly.
\fi
The fact that the 
degree of an
intermediate remainder may be dropped by more than $1$ is taken into account, 
so that
Algorithm~\ref{alg:naivGcdnkernel} works correctly
even if the {\sf GCD} is computed before $n+m-2$ kernel calls.
After $n+m-2$ division steps, either $a$ or $b$ 
becomes zero or constant.
Algorithm~\ref{alg:naivegcd} returns the 
other polynomial as {\sf GCD}.
Figure~\ref{fig:gcdnai} shows one Euclidean division step within a thread-block
after running Algorithm~\ref{alg:naivGcdnkernel} once.

\iflongversion
{
We denote by $W_1$, $S_1$ and $O_1$, the work, span and overhead of Algorithm~\ref{alg:naivegcd}, respectively. Each thread-block performs $2\,{\ell}+1$ arithmetic operations, and each thread makes at most $5$ accesses to the global memory. We obtain the following estimates, where ${\delta}$ stands for $m+n-2$ and ${\mu}$ stands for $n+\ell+1$,
\begin{equation*}
\label{eq:workspanoverheadnaieuclidean}
W_1 = \frac{m\,(2\,n\,{\ell}+{\mu}-2)}{\ell}, \,
S_1 = 3\,{\delta} \ \ {\rm and} \ \
O_1 = \frac{5\,m\,U\,{\mu}}{\ell}.
\end{equation*}

In order to apply Theorem~\ref{thrm:GrahamBrendt},
we shall compute the quantities $N({\cal P})$, $L({\cal P})$
and $C({\cal P})$ defined in Section~\ref{sect:MMM}.
We denote them by $N_{\rm nai}$, $L_{\rm nai}$ and $C_{\rm nai}$, respectively.
One can easily check that we have
\begin{equation*}
\label{eq:brentnaieuclidean}
N_1 = \frac{m\,{\mu}}{\ell}, \,
L_1 = {\delta} \ \ {\rm and} \ \
C_1 = 3+5\,U.
\end{equation*}
}
\else
{
We denote by ${\ell}$ the number of threads per thread-block.
We compute the work, span and overhead, respectively as
$W_1 = \frac{m\,(2\,n\,{\ell}+n+\ell-1)}{\ell}$,
$S_1 = 3\,(m+n-2)$ and
$O_1 = \frac{5\,m\,U\,(n+\ell+1)}{\ell}$.
To apply Corollary~\ref{coro:Ning}, one can easily check that
those three quantities are
$N_1 = \frac{m\,(n+\ell+1)}{\ell}$,
$L_1 = m+n-2$ and
$C_1 = 3+5\,U$.
}
\fi

\begin{algorithm}[htb]
\caption{{\sf NaivePlainGcdGPU($a,b$)}}
\label{alg:naivegcd}

\Indm
\KwIn{  $a, b \in {\KK}[X]$ with ${\deg}(a) \geq {\deg}(b)$ that is, $n-1 \geq m-1$.} 
\KwOut{ $g \in {\KK}[X]$, s.t. $g=${\sf GCD}$(a,b)$.}


\Indp
{\bf int }{ }{\sf st}$[]= \{\deg(a),\deg(b)\}$\;
Let $\ell$ be the number of threads in a thread block\;


\For{$(i= 0; i < n+m-2; i = i+1 )$}
{      {\sf NaivePlainGcdKernel$\lll \lceil  m/\ell \rceil , \ell\ggg (a, b, $ {\sf st})}\; 
}

\eIf{$a$ is a zero or constant  polynomial}
{
Compute $k_b$ the maximum $i$ s.t. $b[i] \neq 0$ holds\;
Let $g$ be array of size $k_b+1$ with coefficients in {\KK} s.t. $g[i] = b[i]$ for $0 \leq i \leq k_b$\; 
}
{
Compute $k_a$ the maximum $i$ s.t. $a[i] \neq 0$ holds\;
Let $g$ be array of size $k_a+1$ with coefficients in {\KK} s.t. $g[i] = a[i]$ for $0 \leq i \leq k_a$\;
}
 {\bf return} $g$\;

\end{algorithm}

\begin{algorithm}[htb]
\caption{{\sf NaivePlainGcdKernel}($a,b,$ {\sf st}   )}
\label{alg:naivGcdnkernel}

\Indm
\KwIn { $a, b \in {\KK}[X]$ and  {\sf st}$[] $ stores the current degree of $a$ and $b$.}
\Indp
$j = ${\sf blockID}$\cdot${\sf blockDim} $ +$ {\sf threadID}\; 

\If{ {\sf st}$[0] \geq$ {\sf st}$[1] > 0 \wedge j <  $  {\sf st}$[1]$ }
{
	$k = j +${\sf st}$[0] -${\sf st}$[1]$\;
	$a[k] = a[k]-b[j] \cdot a[${\sf st}$[0]] \cdot b[${\sf st}$[1]]^{-1}$\;
	\If {$j == $ {\sf st}$[1]-1$}
	{	\While{  $({\sf st}[0] \geq 0) \wedge (a[ {\sf st}[0 ]]= 0)$}
		{		{\sf st}$[0 ] = $ {\sf st}$[0 ]-1$\;
		}
	}
}
\ElseIf{$0 < $ {\sf st}$[0] < $ {\sf st}$[1] \wedge j <  $  {\sf st}$[0]$ }
{
	$k = j +${\sf st}$[1] -${\sf st}$[0]$\;
	$b[k] = b[k]-a[j] \cdot b[${\sf st}$[1]] \cdot a[${\sf st}$[0]]^{-1}$\;

	\If {$j == $ {\sf st}$[0]-1$}
	{	\While{ $({\sf st}[1 ] \geq 0) \wedge (b[ {\sf st}[1 ]]= 0)$}
		{	{\sf st}$[1 ] = $ {\sf st}$[1 ]-1$\;
		}
	}
}

\end{algorithm}

\subsection{Optimized algorithm}
\label{sect:optgcd}

\begin{figure}[htb]
	\centering
	\includegraphics[scale=0.25]{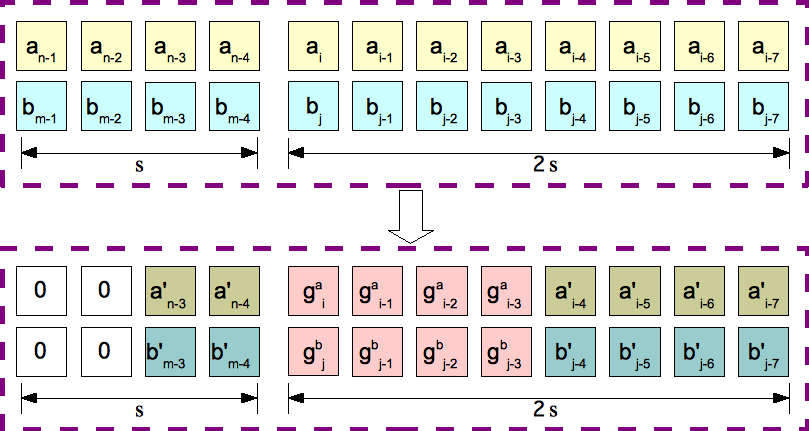}
  \caption{Optimized Euclidean: illustration of a thread-block
  reading coefficients from $a$,$b$ and writing to
 $g^a$,$g^b$.}
  \label{fig:gcdopt}
  \end{figure}

In each kernel, thread-blocks collectively perform 
at most $s$ division steps, instead of one, and update $s$
coefficients from both $a$ and $b$.
After a division step, the degree of the dividend polynomial is decreased
by at least one, and then in the next division step, coefficients from 
the divisor polynomial are adjusted by one or more shift operations.
Thus,  we need  $2\,s$ coefficients 
from both $a$ and $b$ to be sure
that after $s$ division steps we correctly
have $s$ coefficients of both $a$ and $b$.
Since the consecutive thread-blocks 
have $s$ common coefficients of both $a$ and $b$,
the number of thread-blocks is
$\min(\frac{\deg(a)+1}{s}, \frac{\deg(b)+1}{s})$. 
A thread-block also needs $s$-{\em head} coefficients
of both $a$ and $b$ to run $s$ division steps.
Thus, each thread-block uses $3\,s$ threads 
(like Algorithm~\ref{alg:optdivisionkernel}).
Algorithm~\ref{alg:optimizedgcd}
is our top level optimized algorithm
for computing {\sf GCD}($a$, $b$).
\iflongversion
Given a positive integer $s$,
the algorithm calls a kernel
iteratively with
 $\lceil \frac{\deg(b)+1}{s} \rceil$ thread-blocks,
each of them uses $3\,s$ threads.
\fi
Figure~\ref{fig:gcdopt} shows $s$ Euclidean division steps within a thread-block
after running the kernel Algorithm~\ref{alg:optgcdkernel} once.

\iflongversion
Each thread-block, $i$ copies $s$-{\em head}
coefficients from both $a$ and 
$b$. 
It also copies $2\,s$ other coefficients from both $a$ ($b$),
say $X^{d-i\,s}, X^{d-i\,s-1},\cdots, X^{d-i\,s-2s}$,
where $d$ is the degree of $a$ (resp. $b$).
The first $s$ threads in a thread-block
completes $s$ division steps 
with respect to these $s$-{\em head}.
 $u$ and $v$ keeping
 track of current leading coefficients 
of $a$ and $b$ respectively. So, in a division step,
some threads out of these $s$ threads are not active.
 These $s$ threads do not need to write back 
any coefficient. The purpose of those
coefficients and those threads are
to broadcast the leading coefficient
of $a$ and $b$ to other threads only.
Once the first $s$ threads compute the current leading
coefficient
of both $a$ and $b$, the other $2s$ threads complete a 
division step for the other coefficients.
After $s$ division steps 
each thread-block can write back $s$
coefficients correctly to the global memory.
\fi

\iflongversion
{
We denote by $W_s$, $S_s$ and $O_s$, the work, span and overhead respectively. Each thread-block performs $s+2\,(s-1)+\cdots+2\,\frac{s}{2} = \frac{27}{4}\,s^2 + \frac{13}{2}\,s$ arithmetic operations regarding the $s$-head coefficients. Each thread makes at most $8$ accesses to the global memory. We obtain the following estimates, where ${\mu}$ stands for ${\frac {345}{16}}\,{s}^{2}+{\frac {77}{4}}\,s$ and ${\nu}$ stands for $\frac{9}{4}+\frac{6}{s}$,
\begin{equation}
\label{eq:workspanoverheadopteuclidean}
\begin{split}
W_s = {\nu}\,{m}^{2}+ \left( \frac{9}{2}\,n+{\frac {n}{2\,s}}
+{\frac {87}{8}}\,s+\frac{23}{2} \right) m- {\mu}, \\
S_s = 3\,n+3\,m \ \ {\rm and} \ \
O_s = \frac{8\,m\,U\,(n+s)}{s^2}.
\end{split}
\end{equation}

In order to apply Corollary~\ref{coro:Ning},
we shall compute the quantities $N({\cal P})$, $L({\cal P})$
and $C({\cal P})$ defined in Section~\ref{sect:MMM}.
We denote them here by $N_s$, $L_s$ and $C_s$, respectively.
One can easily check that we have
\begin{equation}
\label{eq:brentopteuclidean}
N_s = \frac{m\,n}{s^2}+\frac{m}{s}, \,
L_s = \frac{n}{s}+\frac{m}{s} \ \ {\rm and} \ \
C_s = 3\,s+8\,U.
\end{equation}
}
\else
{
We compute the work, span and overhead, respectively as 
$W_s = (\frac{9}{4}+\frac{6}{s})\,{m}^{2}+ \left( \frac{9}{2}\,n+{\frac {n}{2\,s}}
+{\frac {87}{8}}\,s+\frac{23}{2} \right) m-{\frac {345}{16}}\,{s}^{2}-{\frac {77}{4}}\,s$,
$S_s = 3\,n+3\,m$ and
$O_s = \frac{8\,m\,U\,(n+s)}{s^2}$.
To apply Corollary~\ref{coro:Ning},
one can easily check that those three quantities are
$N_s = \frac{m\,n}{s^2}+\frac{m}{s}$,
$L_s = \frac{n}{s}+\frac{m}{s}$ and
$C_s = 3\,s+8\,U$.
}
\fi

\begin{algorithm}[htb]{}
\caption{{\sf OptimizedPlainGcdGPU($a,b, s$)}}
\label{alg:optimizedgcd}
\Indm
\KwIn{  $a, b \in {\KK}[X]$ with ${\deg}(a) \geq {\deg}(b)$ that is, $n-1 \geq m-1$ and an integer $s>1$.}
\KwOut{ $g \in {\KK}[X]$, s.t. $g=${\sf GCD}$(a,b)$.}

\Indp
{\bf int }{ }{\sf st}$[2]= \{\deg(a),\deg(b)\}$\;
Let $\ell=3s$ be the number of threads in a thread block\;


\For{$(i= 0; i <  {n+m-2} ;i = i + s)$}
{     {\sf OptGcdKer$\lll \lceil  m/s \rceil , \ell \ggg (a, b,s,$ {\sf st})}\;
}

\eIf{$a$ is a zero or constant  polynomial}
{
Compute $k_b$ the maximum $i$ s.t. $b[i] \neq 0$ holds\;
Let $g$ be array of size $k_b+1$ with coefficients in {\KK} s.t. $g[i] = b[i]$ for $0 \leq i \leq k_b$\; 
}
{
Compute $k_a$ the maximum $i$ s.t. $a[i] \neq 0$ holds\;
Let $g$ be array of size $k_a+1$ with coefficients in {\KK} s.t. $g[i] = a[i]$ for $0 \leq i \leq k_a$\;
}
 {\bf return} $g$\;

\end{algorithm}

\begin{algorithm}[htb]{}
\caption{{\sf OptGcdKer($a,b, s$  {\sf st})}}
\label{alg:optgcdkernel}
\Indm
\KwIn {$a, b \in {\KK}[X]$, an integer $s>1$   and 
{\sf st}$[] $ stores the current degree of $a$ and $b$.}

\Indp
Let {\sf sAc, sBc, sA, sB}  be  {\sf local } arrays of size $s,s,2s,2s$  respectively  with coefficients in {\KK}\;

{\sf local} integers $u = v = w = e = 0$\;

$j = ${\sf blockID}$\cdot${\sf blockDim} $ +$ {\sf threadID}; $t = $ {\sf threadID}\;

\tcc{copying from global memory}
\If{$t <s$} {
	{\sf sAc}$[t]= a[${\sf st}$[0]-t]$\;
	{\sf sBc}$[t]= b[${\sf st}$[1]-t]$\;
}

\If{$t\geq s $} {
{\sf sA}$[t-s] = a[${\sf st}$[0]- s\,${\sf blockID}$ - t]$ { } {\sf sB}$[t-s] = b[${\sf st}$[1]- s\,${\sf blockID}$ - t]$\;
}


\For{$(k= 0; k < s; k = k+1)  $}
{
\If{ ({\sf st}$[0] \geq ${\sf st}$[1] \wedge  ${\sf st}$[1] \geq 0$)}
{

	\If{$(u + t < s) \wedge  (v+t < s)$}
	{	{\sf sAc}$[u+t]$ -= {\sf sBc}$[v+t] \cdot${\sf  sAc}$[u] \cdot ${\sf sBc}$[v]^{-1}$\;
		
	}
	\If{$(u+t \geq s) \wedge  (v+t \geq s)$ }
	{
		 {\sf sA}$[w+t-s]$ -= {\sf sB}$[e+t-s]\cdot${\sf  sAc}$[u] \cdot ${\sf sBc}$[v]^{-1}$\;		
	}
	\If{$t == 0$ }
	{	\While{{\sf sAc}$[u]=0$}
		{	$u = u+1$; {  } $w = w+1$; {  }   {\sf st}$[0] = ${\sf st}$[0]-1$\;
		}
	}
}
	\If{ ({\sf st}$[1] \geq ${\sf st}$[0]) \wedge  (${\sf st}$[0] \geq 0)$}
	{	

		\If{$(u + t < s) \wedge   (v+t < s)$}
		{	{\sf sBc}$[v+t]$ -= {\sf sAc}$[u+t] \cdot${\sf  sBc}$[v] \cdot ${\sf sAc}$[u]^{-1}$\;
		}
		\If{$(u + t \geq s) \wedge (v+t \geq s)$}		
		{
			{\sf sB}$[e+t-s]$ -= {\sf sA}$[w+t-s]\cdot${\sf  sBc}$[v] \cdot ${\sf sAc}$[u]^{-1}$\;		

		}

		\If{$t == 0$}
		{	\While{{\sf sBc}$[v]=0$}
			{	$v = v+1$; {  }	$e = e+1$; { } {\sf st}$[1] = ${\sf st}$[1]-1$\;
			}
		}
	}
}

\If{$t\geq s $}
{
  \tcc{writing  to global memory}
	  $a[${\sf st}$[0]- s\,${\sf blockID}$ - t]= $ {\sf sA}$[t-s]$\; 
	   $b[${\sf st}$[1]- s\,${\sf blockID}$ - t] = $ {\sf sB}$[t-s]$ \;
}
\If{$j == \min(${\sf st}$[0]$, {\sf st}$[1] )$}
{	Update {\sf st} array with the new degree of $a$ and $b$\;
	
}
\end{algorithm}

\subsection{Comparison of running time estimates}
\label{sect:analysisgcd}

\iflongversion
{
We first compare the overheads of the two algorithms.
Since we have $2 {\ell} \leq Z$ and $6 s \leq Z$, we 
replace $m$, $s$ and ${\ell}$ by $n$, $Z/6$ and $Z/2$, respectively
in the overhead ratio,
\begin{equation}
\frac{O_1}{O_s} = {\frac {5}{48}}\,{\frac {Z \left( 2\,n+2+Z \right) }{6\,n+Z}}.
\end{equation}
Next, we also observe the ratio $W_1 / W_s$ is asymptotically constant, 
since we have, where $\mu$ stands for $115\,{Z}^{3}+616\,{Z}^{2}$,
\begin{equation}
\frac{W_1}{W_s} = {\frac {(284\,Z+2)\,n^{2}+(Z-2)\,n }{(1296\,Z+7488)\,
{n}^{2}+(348\,{Z}^{2}+2208\,Z)\,n-{\mu}}}
.
\end{equation}
Thus, the improvement has
a fairly low expense.
Applying Theorem~\ref{thrm:GrahamBrendt},
the running times on $p$ 
SMs of the naive and optimized algorithms are 
bounded over by
\begin{equation}
\label{eq:upperboundsdivision3}
(N_1 / p + L_1) \cdot C_1 \ \ {\rm and} \ \ 
(N_s / p + L_s) \cdot C_s.
\end{equation}
When $n$ escapes to infinity, the ratio $R$ is equivalent to
\begin{equation}
\label{eq:upperboundseuclidean2}
\frac{(3+5\,U)\,Z}{3\,(Z+16\,U)}.
\end{equation}
We observe that this ratio is larger than $1$ if and only
if $Z > {\frac {144\,U}{5\,U-6}}$ holds.
Thus, the condition is expected to hold
and, in this case, the optimized algorithm is overall than the naive one.
}
\else
{
Since we have $2 {\ell} \leq Z$ and $6 s \leq Z$, we
replace ${\ell}$ and $s$ by $Z/2$ and $Z/6$, respectively,
and assume that $m$ equals to $n$.
We first observe that the ratio $W_1 / W_s$ is asymptotically constant,
essentially ${\frac {8\,(Z+1)}{3\,(9\,Z+52)}}$,
and the span ratio $S_1 / S_s = \frac{n+m-2}{n+m}$
is asymptotically 1.
Next, we compute the overhead ratio $O_1 / O_s$, that is,
\begin{equation}
\frac{O_1}{O_s} = {\frac {5}{48}}\,{\frac {Z \left( 2\,n+2+Z \right) }{6\,n+Z}}.
\end{equation}
We see that the parallelism overhead 
improvement is done at a fairly low expense in terms of work overhead.
Applying Corollary~\ref{coro:Ning},
we denote the running time estimate ratio $R$ of 
 the naive algorithm over the optimized one, that is,
\begin{equation}
R = {\frac { \left( 6\,n-2+Z \right)  \left( 3+5\,U \right) Z}{ \left( 18
\,n+Z \right)  \left( Z+16\,U \right) }}.
\end{equation}
When $n$ escapes to infinity, the ratio $R$ is equivalent to
\begin{equation}
\label{eq:upperboundseuclidean2}
\frac{(3+5\,U)\,Z}{3\,(Z+16\,U)}.
\end{equation}
We observe that this ratio is larger than $1$ if and only
if $Z > 9.6$ holds.
This condition clearly holds
and the optimized algorithm is overall better.
This is verified in practice~\cite{HM2012}, where $s$ is set to  $256$,
also shown in Table~\ref{tab:gcd}.
}
\fi


\begin{table}[htb]
  \small
  \centering
  \begin{tabular} {| c c | c | c | c | c |}
    \hline
    {$n$} & {$m$} & $s$ = 2 & $s$ = 4 & $s$ = 8 & $s$ = 16 \\
    \hline
    4000 & 4000	& 0.004107	& 0.002775	& 0.003727	& 0.004811 \\
    5000 & 1000	& 0.001684	& 0.001204	& 0.001432	& 0.003645 \\
    5000 & 5000	& 0.008007	& 0.005846	& 0.007808	& 0.011715 \\
    6000 & 1000	& 0.001830	& 0.001298	& 0.001500	& 0.004129 \\
    6000 & 6000	& 0.010359	& 0.007614	& 0.010166	& 0.014352 \\
    7000 & 1000	& 0.001972	& 0.001381	& 0.001614	& 0.004551 \\
    7000 & 7000	& 0.013068	& 0.008821	& 0.012166	& 0.016938 \\
    8000 & 1000	& 0.002111	& 0.001456	& 0.001739	& 0.005015 \\
    8000 & 8000	& 0.016029	& 0.010853	& 0.014877	& 0.019740 \\
    \hline
  \end{tabular}
  \caption{Running time (secs) of the polynomial multiplication algorithm
  with polynomials $a$ ($\deg(a) = n-1$) and $b$ ($\deg(b) = m-1$) 
  and the parameter $s$}
	\label{tab:mul}
\end{table}

\begin{table}[htb]
  \small
  \centering
  \begin{tabular} {| c c | c | c |}
    \hline
    {$n$} & {$m$} & $s$ = 1 & $s$ = 512 \\
    \hline
    2000 & 1500	& 0.058	& 0.024 \\
    3000 & 2500	& 0.108	& 0.039 \\
    4000 & 3500	& 0.158	& 0.053 \\
    5000 & 4500	& 0.203	& 0.069 \\
    6000 & 5000	& 0.235	& 0.056 \\
    7000 & 6000	& 0.282	& 0.066 \\
    8000 & 7000	& 0.324	& 0.076 \\
    9000 & 8000	& 0.367	& 0.087 \\
    10000 & 9000	& 0.411	& 0.097 \\
    \hline
  \end{tabular}
  \caption{Running time (secs) of the Euclidean algorithm for polynomials 
  $a$ ($\deg(a) = n-1$) and $b$ ($\deg(b) = m-1$) 
  with the parameter $s$}
	\label{tab:gcd}
\end{table}

\section{Conclusion}

We have presented a model
of multithreaded computation combining
the fork-join and SIMD parallelisms, with an emphasis
on estimating parallelism overheads, 
so as to reduce communication and synchronization costs 
in GPU programs.

Four applications illustrated the effectiveness of our model.
In each case, we determined a range of values for a program parameter
in order to optimize the corresponding  algorithm 
in terms of parallelism overheads. Experimentation validated
the model prediction.

Our order of magnitude estimates for the program parameter
of radix sort~\cite{SHG2009} agrees with the empirical results
of that paper.

For the Euclidean algorithm, our running time estimates
match those obtained 
with the Systolic VLSI Array Model~\cite{DBLP:journals/tc/BrentK84}.
Moreover, our CUDA code~\cite{HM2012} implementing 
this optimized Euclidean algorithm runs in linear time
w.r.t to the input polynomials degree, up to degree 10,000.

For polynomial multiplication,  our theoretical analysis implies that 
the program parameter $s$ must be as small as possible.
In practice, we could vary this parameter between $2$ and $32$ 
and we found that the optimal value was $4$.
Since certain hardware features are not integrated
into the model, we found that the model prediction was also useful
in that case.


\iflongversion
\section*{Acknowledgment}
The authors would like to thank NSERC of Canada and
Maplesoft Inc. for supporting their work.

They are also grateful to Changbo Chen for his help
in improving earlier versions of this paper.
\fi

\bibliographystyle{abbrv}
\bibliography{reference}

\end{document}